\documentclass[twocolumn,showpacs,preprintnumbers,amsmath,amssymb,pra,aps,
    superscriptaddress,longbibliography]{revtex4-1}
\usepackage[colorlinks=true, linkcolor=gray, citecolor=gray]{hyperref}

\usepackage{graphicx}
\usepackage{amsmath}
\usepackage{amssymb}
\usepackage{mathtools}
\usepackage{soul}
\usepackage{siunitx}
\usepackage[space]{grffile}
\usepackage{tikz}
\usepackage{booktabs}
\usepackage[globalcitecopy]{bibunits}

\newcommand{\ket}[1]{\left\lvert #1 \right\rangle}
\newcommand{\bra}[1]{\left\langle #1 \right\rvert}
\newcommand{\Jtwo}{J_{2}}
\newcommand{\Dzt}{\Delta_{02}}
\newcommand{\UCP}{U_{\mathrm{CP}}}
\newcommand{\UA}{U_{\mathrm{A}}}
\newcommand{\UmA}{U_{\mathrm{-A}}}
\newcommand{\UB}{U_{\mathrm{\phi}}}
\newcommand{\QH}{\mathrm{Q}_{\mathrm{H}}}
\newcommand{\QMone}{\mathrm{Q}_{\mathrm{M1}}}
\newcommand{\QMtwo}{\mathrm{Q}_{\mathrm{M2}}}
\newcommand{\QL}{\mathrm{Q}_{\mathrm{L}}}
\newcommand{\Loneexp}{\tilde L_1}
\newcommand{\Lone}{L_1}
\newcommand{\Li}{L_{1,\mathrm{Idle}}}
\newcommand{\phitwoq}{\phi_{\mathrm{2Q}}}
\newcommand{\phizo}{\phi_{01}}
\newcommand{\phioz}{\phi_{10}}
\newcommand{\phioo}{\phi_{11}}
\newcommand{\phizt}{\phi_{02}}
\newcommand{\phia}{\phi_{\mathrm{a}}}
\newcommand{\phib}{\phi_{\mathrm{b}}}
\newcommand{\phic}{\phi_{\mathrm{c}}}
\newcommand{\phid}{\phi_{\mathrm{d}}}
\newcommand{\degrees}{^{\circ}}

\newcommand{\ns}{\mathrm{ns}}
\newcommand{\us}{\mu\mathrm{s}}
\newcommand{\MHz}{\mathrm{MHz}}
\newcommand{\GHz}{\mathrm{GHz}}
\newcommand{\hours}{\mathrm{h}}
\newcommand{\minutes}{\mathrm{min}}
\newcommand{\Hz}{\mathrm{Hz}}
\newcommand{\Phio}{\Phi_0}
\newcommand{\tp}{t_{\mathrm{p}}}
\newcommand{\toneq}{t_{\mathrm{1Q}}}
\newcommand{\ttwoq}{t_{\phi}}
\newcommand{\tlim}{t_{\mathrm{lim}}}
\newcommand{\tlimp}{t_{\mathrm{lim}}^{+}}
\newcommand{\ts}{t_{\mathrm{s}}}
\newcommand{\ttotal}{t_{\mathrm{total}}}
\newcommand{\Tone}{T_\mathrm{1}}
\newcommand{\Ttwoecho}{T_\mathrm{2}}
\newcommand{\Ttwostar}{T^*_\mathrm{2}}
\newcommand{\Fid}{F}
\newcommand{\Infid}{\varepsilon}
\newcommand{\Fro}{F_\mathrm{RO}}
\newcommand{\Ntwoq}{N_{\mathrm{2Q}}}

\newcommand{\modulo}{\mathrm{mod}}

\begin{document}
\title{High-fidelity controlled-Z gate with maximal intermediate leakage operating at the speed limit in a superconducting quantum processor}
\author{V.~Negîrneac}
\thanks{These authors contributed equally to this work.}
\affiliation{QuTech, Delft University of Technology, P.O. Box 5046, 2600 GA Delft, The Netherlands}
\affiliation{Instituto Superior Técnico, Lisbon, Portugal}
\author{H.~Ali}
\thanks{These authors contributed equally to this work.}
\affiliation{QuTech, Delft University of Technology, P.O. Box 5046, 2600 GA Delft, The Netherlands}
\affiliation{Kavli Institute of Nanoscience, Delft University of Technology, P.O. Box 5046, 2600 GA Delft, The Netherlands}
\author{N.~Muthusubramanian}
\affiliation{QuTech, Delft University of Technology, P.O. Box 5046, 2600 GA Delft, The Netherlands}
\affiliation{Kavli Institute of Nanoscience, Delft University of Technology, P.O. Box 5046, 2600 GA Delft, The Netherlands}
\author{F.~Battistel}
\affiliation{QuTech, Delft University of Technology, P.O. Box 5046, 2600 GA Delft, The Netherlands}
\author{R.~Sagastizabal}
\affiliation{QuTech, Delft University of Technology, P.O. Box 5046, 2600 GA Delft, The Netherlands}
\affiliation{Kavli Institute of Nanoscience, Delft University of Technology, P.O. Box 5046, 2600 GA Delft, The Netherlands}
\author{M.~S.~Moreira}
\affiliation{QuTech, Delft University of Technology, P.O. Box 5046, 2600 GA Delft, The Netherlands}
\affiliation{Kavli Institute of Nanoscience, Delft University of Technology, P.O. Box 5046, 2600 GA Delft, The Netherlands}
\author{J.~F.~Marques}
\affiliation{QuTech, Delft University of Technology, P.O. Box 5046, 2600 GA Delft, The Netherlands}
\affiliation{Kavli Institute of Nanoscience, Delft University of Technology, P.O. Box 5046, 2600 GA Delft, The Netherlands}
\author{W.~Vlothuizen}
\affiliation{QuTech, Delft University of Technology, P.O. Box 5046, 2600 GA Delft, The Netherlands}
\affiliation{Netherlands Organisation for Applied Scientific Research (TNO), P.O. Box 96864, 2509 JG The Hague, The Netherlands}
\author{M.~Beekman}
\affiliation{QuTech, Delft University of Technology, P.O. Box 5046, 2600 GA Delft, The Netherlands}
\affiliation{Netherlands Organisation for Applied Scientific Research (TNO), P.O. Box 96864, 2509 JG The Hague, The Netherlands}
\author{N.~Haider}
\affiliation{QuTech, Delft University of Technology, P.O. Box 5046, 2600 GA Delft, The Netherlands}
\affiliation{Netherlands Organisation for Applied Scientific Research (TNO), P.O. Box 96864, 2509 JG The Hague, The Netherlands}
\author{A.~Bruno}
\affiliation{QuTech, Delft University of Technology, P.O. Box 5046, 2600 GA Delft, The Netherlands}
\affiliation{Kavli Institute of Nanoscience, Delft University of Technology, P.O. Box 5046, 2600 GA Delft, The Netherlands}
\author{L.~DiCarlo}
\affiliation{QuTech, Delft University of Technology, P.O. Box 5046, 2600 GA Delft, The Netherlands}
\affiliation{Kavli Institute of Nanoscience, Delft University of Technology, P.O. Box 5046, 2600 GA Delft, The Netherlands}
\date{\today}

\begin{abstract}
We introduce the sudden variant (SNZ) of the Net Zero scheme realizing controlled-$Z$ (CZ) gates by baseband flux control of transmon frequency.
SNZ CZ gates operate at the speed limit of transverse coupling between computational and non-computational states by maximizing intermediate leakage.
The key advantage of SNZ is tuneup simplicity, owing to the regular structure of conditional phase and leakage as a function of two control parameters.
We realize SNZ CZ gates in a multi-transmon processor, achieving $99.93\pm0.24\%$ fidelity and $0.10\pm0.02\%$ leakage.
SNZ is compatible with scalable schemes for quantum error correction and adaptable to generalized conditional-phase gates useful in intermediate-scale applications.
\end{abstract}
\maketitle

\begin{bibunit}[apsrev4-1]

Superconducting quantum processors~\cite{Kjaergaard19} have recently reached important milestones for quantum computing, notably the demonstration of quantum supremacy on a 53-transmon processor~\cite{Arute19}. On the path to quantum error correction (QEC) and fault tolerance~\cite{Fowler12}, recent experiments have used repetitive parity measurements to stabilize two-qubit entanglement~\cite{Bultink20, Andersen19} and to perform surface-code quantum error detection in a 7-transmon processor~\cite{Andersen20}. These developments have relied on two-qubit controlled-phase (CP) gates realized by dynamical flux control of transmon frequency, harnessing the transverse coupling $\Jtwo$ between a computational state $\ket{11}$ and a non-computational state such as $\ket{02}$~\cite{Strauch03, DiCarlo09}. Compared to other implementations, e.g., cross-resonance using microwave-frequency pulses ~\cite{Sheldon16b} and parametric radio-frequency pulsing~\cite{Hong19}, baseband flux pulses achieve the fastest controlled-$Z$ (CZ) gates (a special case of CP), operating near the speed limit $\tlim=\pi/\Jtwo$~\cite{Barends19}.

Over the last decade, baseband flux pulsing for two-qubit gating has evolved in a continuous effort to increase gate fidelity and to reduce leakage and residual $ZZ$ coupling. In particular, leakage has become a main focus for its negative impact on QEC, adding complexity to error-decoder design~\cite{Varbanov20} and requiring hardware and operational overhead to seep~\cite{Aliferis07,Ghosh13_B,Fowler13,Suchara15,Ghosh15}.
To reduce leakage from linear-dynamical distortion in flux-control lines and limited time resolution in arbitrary waveform generators (AWGs), unipolar square pulses~\cite{DiCarlo09,DiCarlo10} have been superseded by softened unipolar pulses~\cite{Barends14, Kelly15} based on fast-adiabatic theory~\cite{Martinis14}. In parallel, coupling strengths have reduced roughly fourfold (to $\Jtwo/2\pi \sim 10\mathrm{-}20~\MHz$) to reduce residual $ZZ$ coupling, which affects single-qubit gates and idling at bias points, and produces crosstalk from spectator qubits~\cite{Krinner20}. Many groups are actively developing tunable coupling schemes to suppress residual coupling without incurring slowdown~\cite{Chen14,Yan18,Mundada19,Collodo20,Xu20}.

A main limitation to the fidelity of flux-based CP gates is dephasing from flux noise, as one qubit is displaced $0.5\mathrm{-}1~\GHz$ below its flux-symmetry point (i.e., sweetspot~\cite{Schreier08}) to reach the $\ket{11}$-$\ket{02}$ resonance. To address this limitation, Ref.~\onlinecite{Rol19} recently introduced a bipolar variant [termed Net Zero (NZ)] of the fast-adiabatic scheme, which provides a built-in echo reducing the impact of low-frequency flux noise. The double use of the transverse interaction also reduces leakage by destructive interference, as understood by analogy with a Mach-Zehnder interferometer (MZI). Finally, the zero-average characteristic avoids the buildup of long-timescale distortions remaining in flux-control lines after compensation, significantly improving gate repeatability. NZ pulsing has been successfully used in several recent experiments~\cite{Andersen20, Bultink20, Kjaergaard20}, elevating the state of the art for CZ gate fidelity in a multi-transmon processor to $99.72\pm0.35\%$~\cite{Kjaergaard19}. However, NZ suffers from complicated tuneup, owing to the complex dependence of conditional phase and leakage on fast-adiabatic pulse parameters. This complication limits the use of NZ for two-qubit gating as quantum processors grow in qubit count.

In this Letter, we introduce the sudden variant (SNZ) of the Net Zero scheme implementing CZ gates using baseband flux pulsing.
SNZ offers two key advantages while preserving the built-in echo, destructive leakage interference, and repeatability characteristic of conventional NZ.
First, SNZ operates at the speed limit of transverse coupling by maximizing intermediate leakage to the non-computational state.
The second and main advantage is greatly simplified tuneup: the landscapes of conditional phase and leakage as a function of two pulse parameters have very regular structure and interrelation, easily understood by exact analogy to the MZI.
We realize SNZ CZ gates among four pairs of nearest neighbors in a seven-transmon processor and characterize their performance using two-qubit interleaved randomized benchmarking (2QIRB) with
modifications to quantify leakage~\cite{Magesan12, Wood18}. The highest performance achieved has $99.93\pm 0.24\%$ fidelity with $0.10 \pm 0.02\%$ corresponding leakage.
Using numerical simulation with experimental input parameters, we dissect an error budget finding SNZ to slightly outperform conventional NZ.
SNZ CZ gates are fully compatible with scalable approaches to QEC~\cite{Versluis17}.
The generalization of SNZ to arbitrary CP gates is straightforward and useful for optimization~\cite{Lacroix20}, quantum simulation~\cite{Barends15}, and other noisy intermediate-scale quantum (NISQ) applications~\cite{Preskill18}.

A flux pulse harnessing the $\ket{11}$-$\ket{02}$ interaction implements the unitary
\[
\UCP = \begin{pmatrix}
1 &0            & 0             & 0                                 & 0 \\
0 &e^{i\phizo}  & 0             & 0                                 & 0 \\
0 &0            & e^{i\phioz}   & 0                                 & 0 \\
0 &0            & 0             & \sqrt{1-4\Lone}e^{i\phioo}      & \sqrt{4\Lone}e^{\phi_{02,11}} \\
0 &0            & 0             & \sqrt{4\Lone}e^{i\phi_{11,02}}    & \sqrt{1-4\Lone} e^{i\phizt}\\
\end{pmatrix}
\]
in the $\{ \ket{00},\ket{01},\ket{10}, \ket{11}, \ket{02}\}$ subspace, neglecting decoherence and residual interaction between far off-resonant levels.
Here, $\phizo$ and $\phioz$ are the single-qubit phases, and $\phioo=\phizo+\phioz+\phitwoq$, where $\phitwoq$ is the conditional phase.
Finally, $\Lone$ is the leakage parameter.

In a rotating frame which absorbs the single-qubit phases, the system Hamiltonian is
\begin{align*}
H =& \Dzt(t) \ket{02}\!\bra{02}+ \Jtwo \left( \ket{02}\!\bra{11}+ \ket{11}\!\bra{02} \right),
\end{align*}
where $\Dzt(t)$ is the dynamical detuning between $\ket{02}$ and $\ket{11}$.
Each half of the bipolar NZ pulse implements the unitary
\[
\UA=\UmA=\begin{pmatrix}
\alpha e^{i\phia} & \beta e^{i\phib}\\
\beta e^{i\phic} & \alpha e^{i\phid}\\
\end{pmatrix},
\]
in the $\{ \ket{11}, \ket{02} \}$ subspace, where $\alpha,\beta \in \left[0,1\right]$ satisfy $\alpha^2+\beta^2=1$ and $\phia+\phid=\phib+\phic+\pi~(\modulo~2\pi)$.
In the MZI analogy, this unitary is the action of each (ideally identical) beamsplitter.

In SNZ, each half pulse is a square pulse with amplitude $\pm A$ and duration $\tp/2=\tlim/2$.
SNZ intentionally adds an idling period $\ttwoq$ between the half pulses to perfect the analogy to the MZI (Fig.~1 inset), allowing accrual of relative phase $\phi$ in between the beamsplitters.
The unitary action of this idling is
\[
\UB = \begin{pmatrix}
1 & 0\\
0 &  e^{i\phi}\\
\end{pmatrix}.
\]
An ideal CZ gate, our target here, achieves $\phizo=\phioz=0~(\modulo~2\pi)$, $\phitwoq=\pi~(\modulo~2\pi)$ (phase condition PC), and $\Lone=0$ (leakage condition LC), with arbitrary $\phizt$. Accomplishing both conditions with $\UmA\UB\UA$ requires
\[
\alpha^2 e^{i 2\phia} + \beta^2 e^{i\left(\phib +\phic+ \phi \right)}=-1\ \mathrm{(PC)}
\]
simultaneously with either one of three conditions:
(LC1) $\beta=0$; (LC2) $\phia-\phid-\phi=\pi~(\modulo~2\pi)$; or (LC3) $\alpha=0$.
LC1 (LC3) corresponds to perfect reflection (transmission) at each beamsplitter.
LC2 corresponds to destructive interference at the second beamsplitter of the $\ket{02}$ leakage produced by the first.

\begin{figure}
\includegraphics[width=\columnwidth]{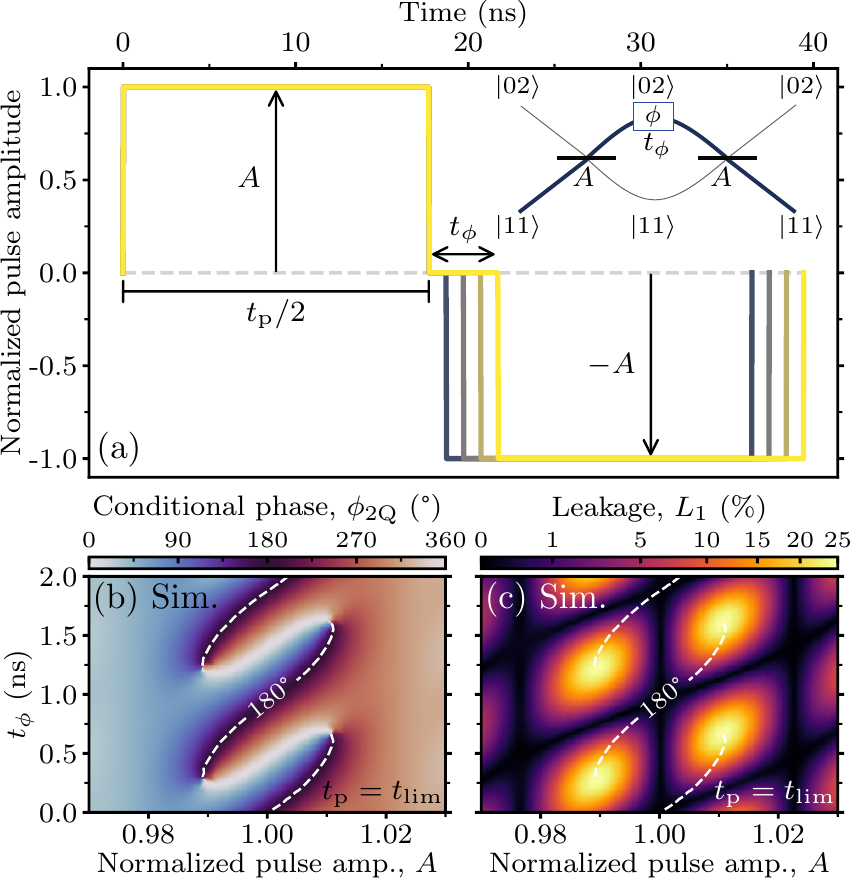}
\caption{
\label{fig:fig1}
Numerical simulation of an idealized SNZ pulse with infinite time resolution for the pair $\QL$-$\QMtwo$, with $\Dzt/2\pi=1.063~\GHz$ (at bias point) and $\tlim=35.40~\ns$ for the $\ket{11}$-$\ket{02}$ interaction. (a) The flux pulse has $\tp=\tlim$ and variable $A$ and $\ttwoq$. Amplitude $A$ is normalized to the $\ket{11}$-$\ket{02}$ resonance. (b, c) Landscapes of conditional phase $\phitwoq$ (b) and leakage $\Lone$ (c) as a function of $A$ and $\ttwoq$. Dashed curves are contours of $\phitwoq=180\degrees$. In (c), the vertical valley at $A=1$ is due to LC3: each half pulse fully transmits $\ket{11}$ to $\ket{02}$, and vice versa. The other two vertical valleys are due to LC1: each half pulse implements a complete (off-resonant) oscillation and thus perfectly reflects.
The diagonal valleys are due to LC2.
Note the simultaneous overlap of $\phitwoq=180\degrees$ contours with the crossing of LC3 and LC2 valleys.
These crossing points occur at $A=1$ and $\ttwoq$ satisfying $\Dzt\ttwoq=0~(\modulo~2\pi)$.
}
\end{figure}

The key advantage of SNZ over conventional NZ is the very straightforward procedure to simultaneously meet PC and LC3.
To appreciate this, consider first the ideal scenario where the pulses can have infinite time resolution. For, $\tp=\tlim$, $\ttwoq=0$ ($\phi=0$), and $A=1$ ($\Dzt=0$) each half pulse implements an iSWAP gate between $\ket{11}$ and $\ket{02}$. Thus, $\alpha=0$ (meeting LC3) and $\phib=\phic=-\pi/2$ (meeting PC). In the MZI analogy, the first beamsplitter fully transmits $\ket{11}$ to $-i\!\ket{02}$ (producing maximal intermediate leakage), and the second fully transmits $-i\!\ket{02}$ to $-\!\ket{11}$.

Consider now the effect of non-zero $\ttwoq$. The idealized two-qutrit numerical simulation with infinite time resolution and no decoherence in Fig.~1 shows that the landscapes of $\phitwoq$ and  $\Lone$ as a function of $A$ and $\ttwoq$ have a clear structure and link to each other. Evidently, $\UmA\UB\UA$ is $2\pi$-periodic in $\phi$, so both landscapes are vertically periodic.
The  $\Lone(A,\ttwoq)$ landscape shows a vertical leakage valley at $A=1$, where LC3 is met.
LC2 gives rise to additional, diagonally running valleys. Juxtaposing the contour of $\phitwoq=180\degrees$ shows that PC is met at the crossing points between these valleys.
In this way, this regular leakage landscape provides useful crosshairs for simultaneously meeting PC.
We note that along the LC3 vertical valley, $\phitwoq$ changes monotonically as a function of $\ttwoq$, allowing the realization of CP gates with any desired $\phitwoq$.
We leave this useful generalization for future work, focusing here on CZ gates.

While in this idealized scenario the idling period is not needed, there are practical reasons to include $\ttwoq$ in experiment: any flux-pulse distortion remaining from the first half pulse during the second (e.g., due to finite pulse rise time) will break the symmetry $\UmA\!=\!\UA$. Due to the time resolution $\ts$ of the AWG used for flux control, $\phi$ can only increment in steps of $-\Dzt\ts$, where $\Dzt$ is the detuning at the bias point. As typically $\Dzt/2\pi=0.5\mathrm{-}1~\GHz$ and $\ts\sim 1~\ns$, one may only use the number of intermediate sampling points for very coarse control. For fine control, we propose to use the amplitude $\pm B$ of the first and last sampling points during $\ttwoq$~\cite{SOM_SNZ}.

With these considerations, we turn to the experimental realization of SNZ CZ gates between the nearest-neighbor pairs among four transmons in a 7-transmon processor. High- and low-frequency transmons ($\QH$ and $\QL$, respectively) connect to two mid-frequency transmons ($\QMone$ and $\QMtwo$) using bus resonators dedicated to each pair. Each transmon has a flux-control line for two-qubit gating, a microwave-drive line for single-qubit gating, and a dispersively coupled resonator with Purcell filter for readout~\cite{Heinsoo18, Bultink20}. All transmons can be measured simultaneously by frequency multiplexing using a common feedline. See~\cite{SOM_SNZ} for device details and a summary of measured transmon parameters, single-qubit-gate and readout performance. Each transmon is biased at its sweetspot using static flux bias to counter residual offsets. Flux pulsing is performed using a Zurich Instruments HDAWG-8 $(\ts=1/2.4~\ns)$. Following prior work~\cite{Rol19, Rol20},  we measure the linear-dynamical distortions in the flux-control lines using the Cryoscope technique and correct them using real-time filters built into the AWG.

\begin{figure}
\includegraphics[width=\columnwidth]{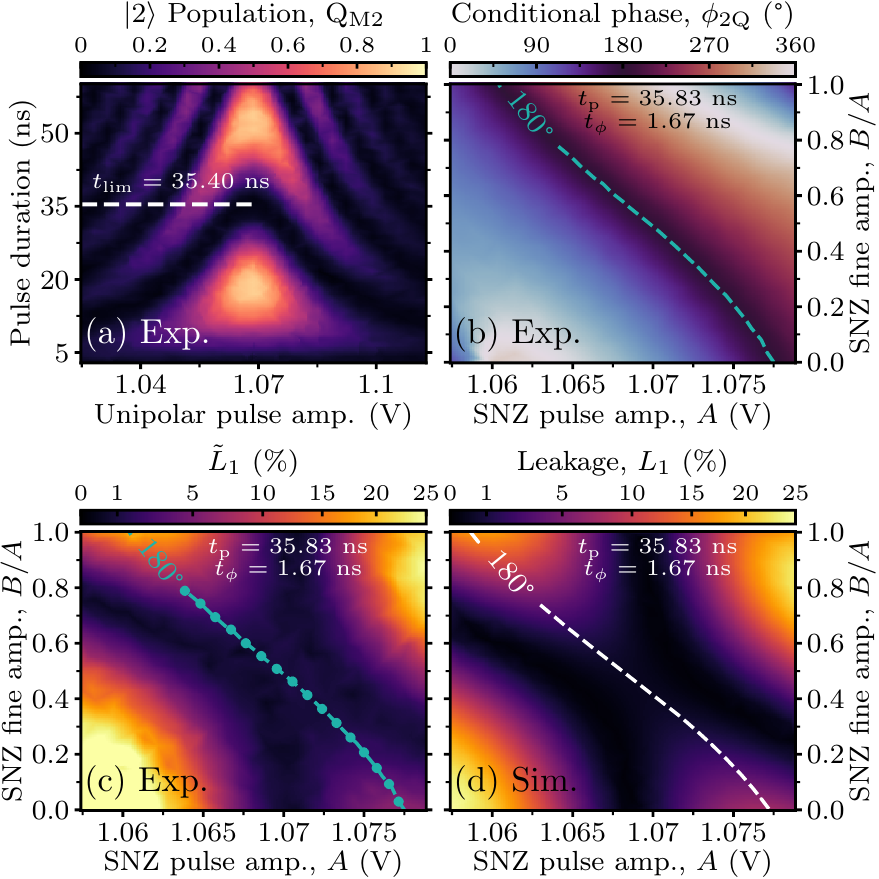}
\caption{
\label{fig:speedlimit}
Calibration of the SNZ pulse for pair $\QL$-$\QMtwo$ and comparison to simulation.
(a) $\ket{2}$-state population of $\QMtwo$ as a function of the amplitude and duration of a unipolar square pulse making $\ket{11}$ interact with $\ket{02}$.
The characteristic chevron pattern is used to identify $\tlim$ and the amplitude bringing $\ket{11}$ and $\ket{02}$ on resonance. (b,c) Landscapes of conditional phase $\phitwoq$ and leakage estimate $\Loneexp$ as a function of SNZ pulse amplitudes $A$ and $B$, with $\tp=\tlimp$ and $\ttwoq=1.67~\ns$. Note that the juxtaposed $\phitwoq=180\degrees$ contour runs along the opposite diagonal compared to Figs.~1(b,c) because increasing $B$ (which decreases $\Dzt$) changes $\phi$ in the opposite direction from $\ttwoq$. Data points marked with dots are measured with extra averaging for detailed examination in Fig.~3.
(d) Numerical simulation of leakage $\Lone$ landscape and $\phitwoq=180\degrees$ contour. The simulation uses pulse and transmon parameters from experiment (neglecting decoherence), and includes the measured final flux-pulse distortion.
All landscapes are sampled using an adaptive algorithm based on~\cite{Nijholt2019}.
}
\end{figure}

We exemplify the tuneup of SNZ pulses using pair $\QL$-$\QMtwo$ (Fig.~2). We first identify $\tlim$ for the $\ket{11}$-$\ket{02}$ interaction and the amplitude $A$ bringing the two levels on resonance.
(The rightmost index indicates the excitation level of the fluxed transmon, here $\QMtwo$). These parameters are extracted from the characteristic chevron pattern of $\ket{2}$-population in $\QMtwo$  as a function of the amplitude and duration of a unipolar square flux pulse acting on $\ket{11}$ [Fig.~2(a)]. The chevron symmetry axis corresponds to $A=1$ and the oscillation period along this axis gives $\tlim$. We set $\tp=\tlimp \equiv 2n\ts$, where $n$ is the smallest integer satisfying $2n\ts \geq \tlim$. Next, we measure the landscapes of $\phitwoq$ and leakage estimate~$\Loneexp$ in the range $A\in\left[0.9, 1.1\right]$, $B\in\left[0,A\right]$. These quantities are extracted from the conditional-oscillation experiments described in~\cite{Rol19}. As expected, the landscape of $\Loneexp$ [Fig.~2(c)] reveals  a vertical valley at $A=1$ and a diagonal valley. Juxtaposing the $\phitwoq=180\degrees$ contour extracted from the $\phitwoq$ landscape [Fig.~2(b)], we observe the matching of PC at the crossing of these valleys. These experimental observations are in excellent agreement with a numerical two-qutrit simulation [Fig.~2(d)].

\begin{figure}
\includegraphics[width=\columnwidth]{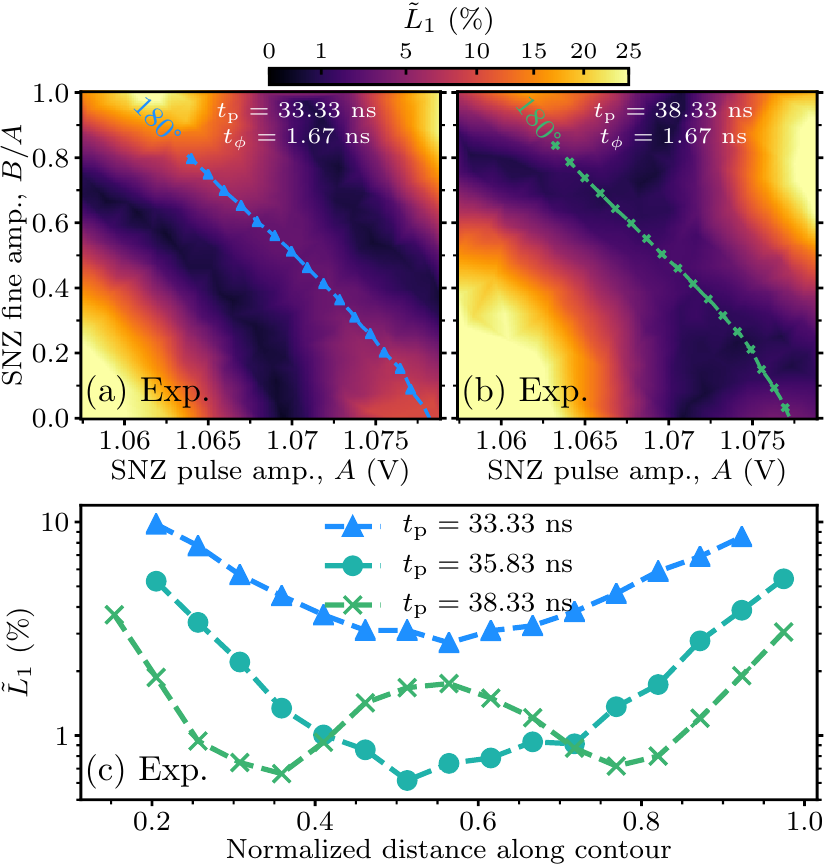}
\caption{\label{fig:too_short_too_long}
Landscapes of the leakage estimate $\Loneexp$ for (a) intentionally short $(\tp=\tlimp-6 \ts)$ and (b) intentionally long $(\tp=\tlimp+6\ts)$ flux pulses on $\QMtwo$.
These landscapes are sampled using an adaptive algorithm based on~\cite{Nijholt2019}.
(c) Extracted $\Loneexp$ along the $\phitwoq=180\degrees$ contours from (a), (b), and Fig.~2(c). For $\tp<\tlim$, a single minimum is observed, but at a higher value than the minimum for $\tp=\tlimp$. For $\tp>\tlim$, two leakage minima are found, matching the single minimum for $\tp=\tlimp$.
}
\end{figure}

Experimentally, it is nearly impossible to precisely match $\tp=\tlim$ due to the discrete $\ts$. To understand the possible consequences of $\tp$ mismatch, we examine the $\phitwoq$ and $\Loneexp$ landscapes for SNZ pulses with intentionally set $\tp = \tlimp \pm 6 \ts$ (Fig.~3). We find that the PC contour remains roughly unchanged in both cases. However, there are significant effects on $\Loneexp$. In both cases, we observe that $\Loneexp$ lifts at the prior crossing of LC2 and LC3 valleys where $\phitwoq=180\degrees$. For too-short pulses [Fig.~3(a)], there remain two valleys of minimal $\Loneexp$, but these are now curved and do not cross the $\phitwoq=180\degrees$ contour. For too-long pulses [Fig.~3(b)], there are also two curved valleys. Crucially, these cross the $\phitwoq=180\degrees$ contour, and it remains possible to achieve PC and minimize leakage at two $(A,B)$ settings. Extracting $\Loneexp$ along the $\phitwoq=180\degrees$ contours  [Fig.~3(c)] confirms that the minimal leakage obtainable for $\tp = \tlimp + 6\ts$ matches that for $\tp=\tlimp$.  The observed impossibility to achieve minimal leakage at $\phitwoq=180\degrees$ for $\tp<\tlim$ is a clear manifestation of the speed limit set by $\Jtwo$. In turn, the demonstrated possibility to do so for $\tp>\tlim$ (even when overshooting by several sampling points) is an important proof of the viability of the SNZ pulse in practice.

With these insights, we proceed to tuning SNZ CZ gates for the four transmon pairs, following similar procedures.
Namely, we use final weak bipolar pulses of total duration $\toneq=10~\ns$ to null the single-qubit phases in the frame of microwave drives. Also, since our codeword-based control electronics has a $20~\ns$ timing grid, and
$40~\ns<\ttotal=\tp+\ttwoq+\toneq<60~\ns$ for all pairs,  we allocate $60~\ns$ to every CZ gate. However, some  pair-specific details must be noted.  Owing to the overlap of qubit frequencies between mid-frequency qubits, implementing CZ between $\QH$ and $\QMone$ $(\QMtwo)$ requires parking of $\QMtwo$ $(\QMone)$ during the SNZ pulse on $\QH$~\cite{Versluis17,Andersen20}. The parking flux pulse is also bipolar, with each half a square pulse lasting $(\tp+\ttwoq)/2$. Its amplitude is chosen to downshift the parked qubit by $\sim 300~\MHz$, and fine tuned to null its single-qubit phase. For most pairs, we employ the $\ket{11}$-$\ket{02}$ interaction, which requires the smallest flux amplitude (reducing the impact of dephasing from flux noise) and does not require crossing any other interaction on the way to and from it. However, for pair $\QL$-$\QMone$, we cannot reliably use this interaction as there is a flickering two-level system (TLS) aligned with the $\QMone$ qubit transition at this amplitude (See~\cite{SOM_SNZ} for chevron measurements showing the flickering nature of this TLS). For this pair, we therefore employ the $\ket{11}$-$\ket{20}$ interaction. Using square pulses is a side benefit of SNZ in this case: it minimizes exchange between $\ket{01}$ and the TLS, $\ket{11}$ and $\ket{20}$, and  $\ket{01}$ and $\ket{10}$ as their resonances are crossed as suddenly as possible.

\begin{table}
{\footnotesize
\begin{tabular}{lcccc}
\hline
Parameter & $\QMone$-$\QH$  & $\QMtwo$-$\QH$   &  $\QL$-$\QMone$   &  $\QL$-$\QMtwo$  \\
\hline
$\tlim~(\ns)$  & $32.20$ & $29.00$ & $40.60$ & $35.40$ \\
$\tp~(\ns)$  & $32.50$ & $29.16$ & $40.83$ & $35.83$ \\
$\ttwoq~(\ns)$  & 2.92 & 3.75 & 1.25 & 1.67 \\
$(\tp+\ttwoq)/\tlim$ & $1.10$ & $1.13$ & $1.04$ & $1.06$ \\
$\ttotal~(\ns)$ & $45.42$ & $42.91$ & $52.08$ & $47.50$ \\
Interaction & $\ket{11}$-$\ket{02}$ & $\ket{11}$-$\ket{02}$ & $\ket{11}$-$\ket{20}$ & $\ket{11}$-$\ket{02}$ \\
Parked~qubit & $\QMtwo$  & $\QMone$ & -- & -- \\
Avg.~$\Fid$ $(\%)$ & $98.89\pm0.35$ & $99.54\pm0.27$ & $93.72\pm 2.10$ & $97.14\pm 0.72$ \\
Avg.~$\Lone$ $(\%)$ & $0.13\pm0.02$ & $0.18\pm0.04$ & $0.78\pm0.32$ & $0.63\pm0.11$ \\
Max.~$\Fid$ $(\%)$ & $99.77\pm0.23$  & $99.93\pm0.24$  & $99.15\pm1.20$  & $98.56\pm0.70$ \\
Min.~$\Lone$ $(\%)$ & $0.07\pm0.04$ & $0.10\pm0.02$ & $0.04\pm0.08$ & $0.41\pm0.10$ \\
\hline
\end{tabular}
}
\caption{
Summary of SNZ CZ pulse parameters and achieved performance for the four transmon pairs. All SNZ CZ gates null single-qubit phases with weak bipolar square pulse of duration $\toneq=10~\ns$ immediately following the strong pulse. We allocate $60~\ns$ to every CZ gate to conform to the $20~\ns$ timing grid of our control electronics. Gate fidelities and leakage are obtained from 2QIRB keeping the other two qubits in $\ket{0}$. Statistics (average and standard deviation) are taken from repeated 2QIRB runs (see~\cite{SOM_SNZ} for technical details). The maximum $\Fid$ and minimum $\Lone$ quoted are not necessarily from the same run.
}
\end{table}

Table~1 summarizes the timing parameters and performance attained for the four SNZ CZ gates. The CZ gate fidelity $\Fid$ and leakage $\Lone$ are extracted using a 2QIRB protocol modified to quantify leakage~\cite{Wood18, Rol19}.
For each pair, we report the best, average and standard deviation of both values based on at least 10 repetitions of the protocol spanning more than $8~\hours$~\cite{SOM_SNZ}.
Several observations can be drawn. First, CZ gates involving $\QH$ perform better on average than those involving $\QL$. This is likely due to the shorter $\tlim$ and correspondingly longer time $60~\ns-\tp$ spent near the sweetspot. Another possible reason is that the frequency downshifting required of $\QH$ to interact with $\QMone$ and $\QMtwo$  is roughly half that required of the latter to interact with $\QL$. This reduces the impact of dephasing from flux noise during the pulse. Not surprisingly, performance is worst for the pair $\QL$-$\QMone$. Here, the pulse must downshift $\QMone$ the most in order to reach the distant $\ket{11}$-$\ket{20}$ interaction, increasing dephasing from flux noise. Also, there may be residual exchange with the identified TLS and as the $\ket{11}$-$\ket{02}$ and $\ket{01}$-$\ket{10}$ resonances are crossed. Overall, there is significant variation in the performance metrics during repeated 2QIRB characterization. We believe this reflects the underlying variability of qubit relaxation and dephasing times, which however were not tracked simultaneously. In addition to having the best average performance, pair $\QMtwo$-$\QH$ also displays the hero performance based on a single 2QIRB run (Fig.~4). Peaking at $\Fid=99.93\pm0.24\%$, we believe this is the highest CZ fidelity extracted from 2QIRB characterization in a multi-transmon processor.

\begin{figure}
\includegraphics[width=\linewidth]{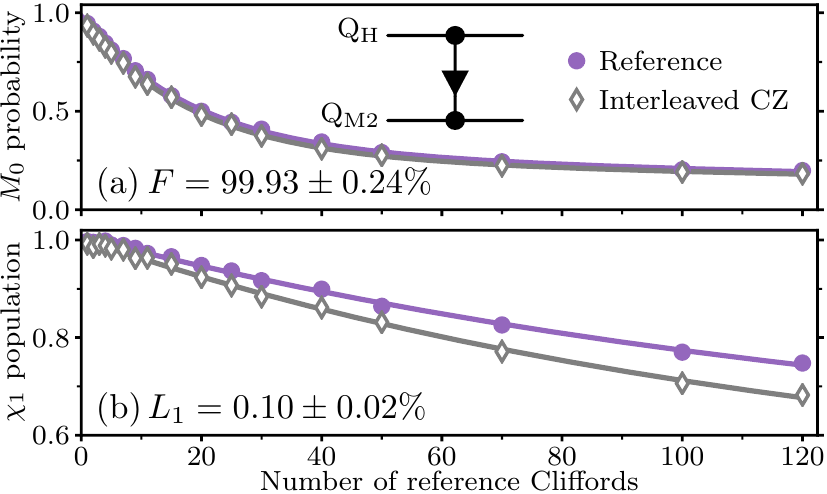}
\caption{
\label{fig:RB}
Best SNZ CZ gate performance achieved from a single run of 2QIRB.
(a) Reference and CZ-interleaved return probability $M_0$ to $\ket{00}$ and (b) population in the computational space $\chi_1$ as a function of the number of two-qubit Cliffords in the reference curve.
Errors bars in $\Fid$ and $\Lone$ are obtained from the uncertainty of exponential-decay fits.
}
\end{figure}

In an effort to identify the dominant sources of infidelity $\Infid=1-\Fid$ and leakage for SNZ CZ gates, we perform a two-qutrit numerical simulation for pair $\QMtwo$-$\QH$ with an error model taking parameters from experiment [Fig.~5]. As in our previous work on conventional NZ~\cite{Rol19}, the simulation incrementally adds: (A) no noise; (B) energy relaxation; (C) Markovian dephasing; (D) dephasing from low-frequency flux noise; and (E) flux-pulse distortion. The experimental inputs for models B, C and D combine measured qubit relaxation time $\Tone$ at the bias point, and measured echo and Ramsey dephasing times ($\Ttwoecho$ and $\Ttwostar$) as a function of qubit frequency. The input to E consists of a final Cryoscope measurement of the flux step response using all real-time filters. The simulation suggests that the main source of $\Infid$ is Markovian dephasing (as in~\cite{Rol19}), while the dominant contribution to $\Lone$ is low-frequency flux noise. The latter contrasts with Ref.~\cite{Rol19}, where simulation identified flux-pulse distortion as the dominant leakage source. We identify two possible reasons for this difference: in the current experiment, the $1/f$ low-frequency flux noise is $\sim\!4$ times larger (in units of $\Phio/\sqrt{\Hz}$) and the achieved flux step response is noticeably sharper. Finally, we use the simulation to compare performance of SNZ to conventional NZ CZ. For the latter, we fix $\ttwoq=0$, $\toneq=60~\ns-\tp$, and use the fast-adiabatic pulse shape and $\tp=45.83~\ns$ optimized by simulation. Overall, the error sources contribute very similarly to the error budget for both cases. The marginally higher overall performance found for SNZ is likely due to the increased time spent at the sweetspot during the $60~\ns$ allocated for each CZ.

\begin{figure}
\includegraphics[width=\columnwidth]{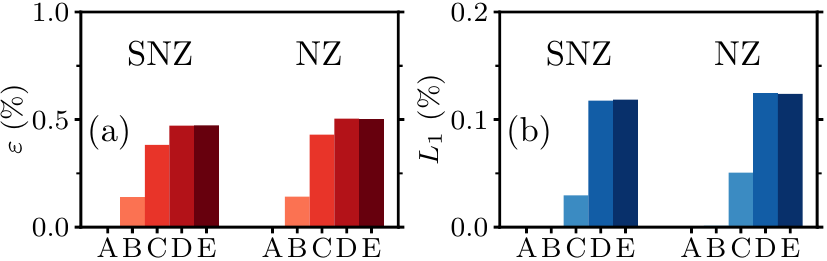}
\caption{
\label{fig:error_budgets}
Error budgets for infidelity $\Infid$ (a) and leakage $\Lone$ (b) obtained by a numerical simulation (as in \cite{Rol19}) of the $\QMtwo$-$\QH$ SNZ CZ gate with parameters in Fig.~4 and for a conventional NZ gate with optimized parameters (see text for details). The simulation incrementally adds errors using experimental input parameters for this pair: (A) no noise; (B) relaxation; (C) Markovian dephasing; (D) dephasing from quasistatic flux noise; and (E) flux-pulse distortion.
}
\end{figure}

In summary, we have proposed and implemented the sudden variant of the NZ pulsing scheme~\cite{Rol19} realizing flux-based CZ gates by exploiting transverse coupling between computational and non-computation states.
SNZ operates at the speed limit of transverse coupling by maximizing the intermediate leakage to the non-computational state.
The key advantage of SNZ over conventional NZ is ease of tuneup, owing to the simple structure of error landscapes as a function of pulse parameters.
We have demonstrated high-fidelity CZ gates between four transmon pairs in a patch of our 7-transmon processor.
To the best of our knowledge, the best fidelity extracted from 2QIRB extends the state of the art.
Control architectures without a timing grid will additionally benefit from the increased speed of SNZ over conventional NZ by reducing the total gate time and thus the impact of decoherence.
Taking advantage of the tuning simplicity, we already employ SNZ CZ gates in the Starmon-5 quantum processor publicly available via the QuTech Quantum Inspire platform~\cite{QuantumInspire}.
Moving forwards, the full compatibility of SNZ with our proposed~\cite{Versluis17} scalable scheme for surface coding makes SNZ our choice for CZ gates as we pursue quantum error correction.
Finally, the noted straightforward extension of SNZ to arbitrary conditional-phase gates will find immediate application in NISQ applications.

\begin{acknowledgments}
We thank L.~Janssen, M.~Rol, M.~Sarsby, T.~Stavenga, and B.~Tarasinski for experimental assistance, C.~Eichler and B.~Terhal for discussions, and G.~Calusine and W.~Oliver for providing the travelling-wave parametric amplifier used in the readout amplification chain. This research is supported by the Office of the Director of National Intelligence (ODNI), Intelligence Advanced Research Projects Activity (IARPA), via the U.S. Army Research Office Grant No. W911NF-16-1-0071, and by Intel Corporation. The views and conclusions contained herein are those of the authors and should not be interpreted as necessarily representing the official policies or endorsements, either expressed or implied, of the ODNI, IARPA, or the U.S. Government. F.~B. is supported by ERC Grant EQEC No.~682726.
\end{acknowledgments}

\end{bibunit}

\begin{bibunit}[apsrev4-1]
\onecolumngrid
\clearpage
\renewcommand{\theequation}{S\arabic{equation}}
\renewcommand{\thefigure}{S\arabic{figure}}
\renewcommand{\thetable}{S\arabic{table}}
\renewcommand{\bibnumfmt}[1]{[S#1]}
\renewcommand{\citenumfont}[1]{S#1}
\setcounter{figure}{0}
\setcounter{equation}{0}
\setcounter{table}{0}

\section*{Supplemental material for 'High-fidelity controlled-Z gate with maximal intermediate leakage operating at the speed limit in a superconducting quantum processor'}
\date{\today}
\maketitle

This supplement provides additional information in support of statements and claims made in the main text.
Section \ref{sec:one} summarizes the main differences between conventional NZ pulses and SNZ pulses.
Section \ref{sec:two} provides further details on the device used and measured transmon parameters.
Section \ref{sec:three} presents the characterization of single-qubit gate performance.
Section \ref{sec:four} provides evidence for the two-level system affecting the realization of SNZ CZ gates in pair $\QL$-$\QMone$ using the $\ket{11}$-$\ket{02}$ interaction.
Section \ref{sec:five} presents a characterization of the residual $ZZ$ coupling between qubits at the bias point.
Section \ref{sec:six} summarizes the technical details of the CZ characterization by repeated 2QIRB.

\section{Comparison of conventional NZ pulses and SNZ pulses}
\label{sec:one}

This section highlights the main differences between conventional NZ pulses and the SNZ pulses introduced here.
The conventional NZ strong pulse [Fig.~S1(a)] consists of two back-to-back half pulses of duration $\tp/2$ each, applied on the higher-frequency transmon. Typically, $\tp/\tlim \sim 1.1\mathrm{-}1.6$. The strong half pulses are formally parametrized as in~\cite{Martinis14}. For the purposes of illustration, here we can loosely lump this parametrization as affecting the amplitude ($\pm A$) and curvature ($A'$) of the strong half pulses. Immediately following the strong pulse, weak bipolar pulses of duration $\toneq$ are applied on both the higher- and lower-frequency transmons with amplitudes $\pm C$ and $\pm D$, respectively, in order to null the single-qubit phases acquired by each. Typically, $\toneq=10~\ns$.
In conventional NZ there is no intermediate idling period between the strong half pulses, so the analogy to the MZI is not exact [Fig.~S1(c)].
During tuneup, one searches the $(A,A')$ space to achieve $\UmA\UA=\UCP(\phitwoq=\pi)$ by only affecting the unitary action of the two beamsplitters.
Because for typical $\tp$ conventional NZ produces significant leakage at the first strong pulse, achieving minimal leakage relies on meeting LC2.
The structure of the $\phitwoq(A,A')$ and $\Lone(A,A')$  landscapes and especially their interrelation are not straightforward, so the search for an $(A,A')$ setting satisfying both PC and LC2 is not easily guided.
We point the interested reader to~\cite{Rol19} for examples.

The SNZ pulses introduced here [Fig.~S1(b)] differ in two key ways. First, the strong half pulses are replaced by square half pulses each with duration $\tp/2$ as close as possible to $\tlim/2$ (as allowed by the AWG sampling period) but not shorter. Second, an intermediate idling period $\ttwoq$ is added to accrue relative phase $\phi$ between $\ket{02}$ and $\ket{11}$, perfecting the analogy to the MZI [Fig.~S1(d)]. We use the amplitude $\pm B$ of the first and last sampling points in $\ttwoq$ and the number of intermediate zero-amplitude points to achieve fine and coarse control of $\phi$, respectively. As in conventional NZ, we use weak bipolar pulses on both transmons (also with $\toneq=10~\ns$) to null the single-qubit phases. During tuneup, we search the $(A,B)$ space to achieve $\UmA\UB\UA=\UCP(\phitwoq=\pi)$. As shown in the main text, the SNZ pulse design gives a very simple structure to the $\phitwoq(A,B)$ and $\Lone(A,B)$ landscapes. Crucially, the crossing point of leakage valleys satisfying LC2 and LC3 matches $\phitwoq=180\degrees$. This simplicity of tuneup is the key advantage of SNZ over conventional NZ.

Another advantage of SNZ over conventional CZ is the reduced total time $\ttotal=\tp+\ttwoq+\toneq$ required to achieve a CZ gate. However, due to the $20~\ns$ timing grid of our control electronics and the transverse coupling strengths
in our device, this speedup is insufficient to reduce the total time allocated per CZ gate from $60$ to $40~\ns$. Nonetheless, in SNZ, the fluxed transmon spends more time at its sweetspot, which reduces the dephasing due to flux noise.

\begin{figure}[h!]
    \includegraphics[width=0.8\columnwidth]{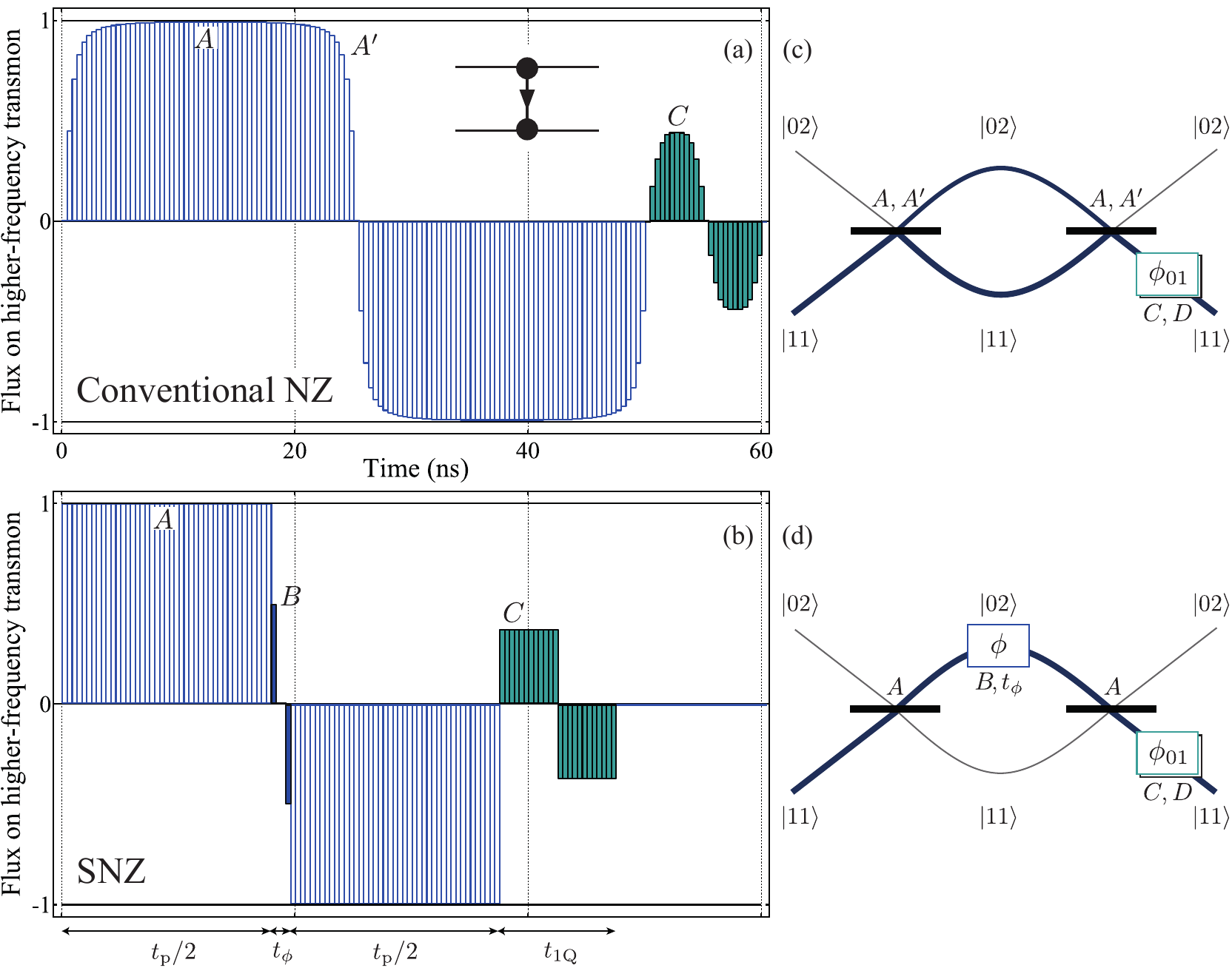}
    \caption{
    Comparison of conventional NZ and SNZ pulses for CZ gates.
    (a) Conventional NZ CZ pulses consist of two back-to-back strong half pulses of duration $\tp/2$ each, followed by two weak back-to-back half pulses of duration $\toneq/2$ each on the higher-frequency qubit. The amplitude ($\pm A$) and curvature ($A'$) of the strong pulses are jointly tuned to set the conditional phasse $\phitwoq$ at minimal leakage $\Lone$ , while the amplitude $\pm C$ of the weak pulses is used to null the single-qubit phase on the higher-frequency transmon. Weak pulses (amplitude $\pm D$) on the lower-frequency qubit (not shown here) are also used to null its single-qubit phase.
    (b) In SNZ, the strong pulses are replaced by square pulses with $\tp$ as close as possible to $\tlim$ but not shorter. Also, an intermediate idling period $\ttwoq$ is added to accrue relative phase $\phi$ between $\ket{02}$ and $\ket{11}$. The amplitude $\pm B$ of the first and last sampling points in $\ttwoq$ and the number of intermediate zero-amplitude points provide fine and coarse control of this relative phase, respectively. SNZ CZ gates also use weak bipolar pulses (now square) of total duration $\toneq$ to null single-qubit phases.
    (c) The MZI analogy for conventional NZ pulses is incomplete. Each strong half pulse implements a beamsplitter (ideally identical) with scattering parameters affected by $A$ and $A'$. However, there is no possibility to independently control the relative phase in the two arms between the beamsplitters.
    (d) The MZI analogy is exact for SNZ pulse. The scattering at the beamsplitters is controlled by $A$ and the relative phase $\phi$ is controlled finely using $B$ and coarsely using $\ttwoq$.
    }
\end{figure}

\section{Device and transmon parameters}
\label{sec:two}
Our experimental study focuses on four transmons in a patch of our 7-qubit processor.
An optical image of the device, zoomed in to these four transmons, is shown in Fig.~S2.
Transmons $\QH$ and $\QL$ both connect to $\QMone$ and $\QMtwo$ with a dedicated coupling bus resonator for each connection.
Every transmon has a dedicated microwave-drive line for single-qubit gating, a flux-control line used for CZ gating, and a dispersively-coupled readout resonator with dedicated Purcell filter~\cite{Heinsoo18, Bultink20} for readout.
Readout is performed by frequency multiplexed measurement of a common feedline capacitively connected to all four Purcell filters. Table~S1 provides a summary of measured parameters for the four transmons.

\begin{figure}[h!]
    \includegraphics[width=0.7\columnwidth]{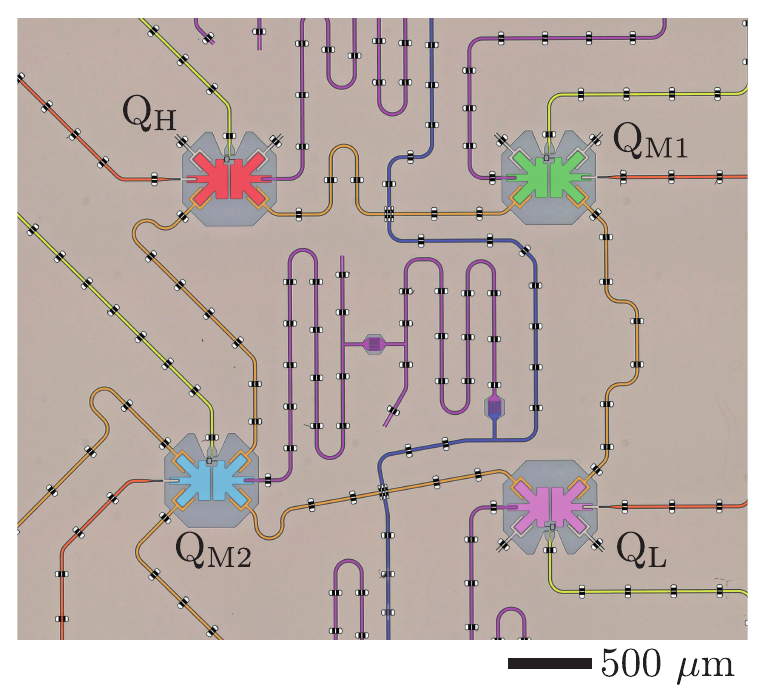}
    \caption{
    Optical image of the device, zoomed in to the four transmons used in this study and with added false color to help identify circuit elements.
    Transmons $\QH$ (red) and $\QL$ (pink) each connect to $\QMone$ (green) and $\QMtwo$ (cyan) using dedicated coupling bus resonators for each pair (light orange).
    Each transmon has a flux-control line for two-qubit gating (yellow), a microwave-drive line for single-qubit gating (dark orange), and dispersively-coupled resonator with Purcell filter for readout (purple)~\cite{Heinsoo18, Bultink20}. The readout-resonator/Purcell-filter pair for $\QMtwo$ is visible at the center of this image.
    The vertically running common feedline (blue) connects to all Purcell filters, enabling simultaneous readout of the four transmons by frequency multiplexing.
    Air-bridge crossovers enable the routing of all input and output lines to the edges of the chip, where they connect to a printed circuit board through aluminum wirebonds.
    }
\end{figure}

\begin{table}[h!]
    \setlength{\tabcolsep}{12pt}
    \renewcommand{\arraystretch}{1.2}
    \begin{tabular}{lcccc}
    \hline
    & $\QH$  & $\QMone$   &  $\QMtwo$  &  $\QL$  \\
    \hline
    Qubit transition frequency at sweetspot, $\omega_\mathrm{q}/2\pi$ $(\GHz)$ & 6.4329 & 5.7707 & 5.8864 & 4.5338 \\
    Transmon anharmonicity, \(\alpha/2\pi\) $(\MHz)$ & -280 & -290 & -285 & -320 \\
    Readout frequency, $\omega_\mathrm{r}/2\pi$ $(\GHz)$ & 7.4925 & 7.2248 & 7.0584 & 6.9132 \\
    Relaxation time, $\Tone$ $(\us)$  & $37\pm1$ & $40\pm1$ & $47\pm1$ & $66\pm1$ \\
    Ramsey dephasing time, $\Ttwostar$ $(\us)$  & $38\pm1$ & $49\pm1$ & $47\pm1$ & $64\pm1$ \\
    Echo dephasing time, $\Ttwoecho$ $(\us)$  & $54\pm2$ & $68\pm1$ & $77\pm1$ & $94\pm2$ \\
    Residual qubit excitation, $(\%)$ & 1.4 & 1.2 & 4.3 & 1.7 \\
    Best readout fidelity, $\Fro$ (\%) & 99.1 & 98.5 & 99.4 & 97.8 \\
    \hline
    \end{tabular}
    \caption{%
    \label{tab:msmt_qb_coh}%
    Summary of frequency, coherence, residual excitation, and readout parameters of the four transmons. The statistics of coherence times for each transmon are obtained from 30 repetitions of standard time-domain measurements~\cite{Krantz19} taken over $\sim\!4~\hours$. The residual excitation is extracted from double-Gaussian fits of single-shot readout histograms with the qubit nominally prepared in $\ket{0}$. The readout fidelity quoted is the average assignment fidelity~\cite{Bultink18}, extracted from single-shot readout histograms after mitigating residual excitation by post-selection on a pre-measurement.
    }
\end{table}

\section{Single-qubit gate performance}
\label{sec:three}
All single-qubit gates are implemented as DRAG-type~\cite{Motzoi09, Chow10b} microwave pulses with a total duration of $4\sigma=20~\ns$, where $\sigma$ is the Gaussian width of the main-quadrature Gaussian pulse envelope.
We perform two sets of experiments to jointly quantify the infidelity $\Infid$ and leakage $\Lone$ of these gates.
First, we perform individual single-qubit randomized benchmarking (1QRB) keeping the other three qubits in $\ket{0}$.
Second, we perform simultaneous single-qubit randomized benchmarking (S1QRB) on pairs of qubits, keeping the other two qubits in $\ket{0}$.
The results obtained from both types of experiment are reported as diagonal and off-diagonal elements in the matrices presented in Fig.~S3.

\begin{figure}[h]
    \includegraphics[width=0.7\columnwidth]{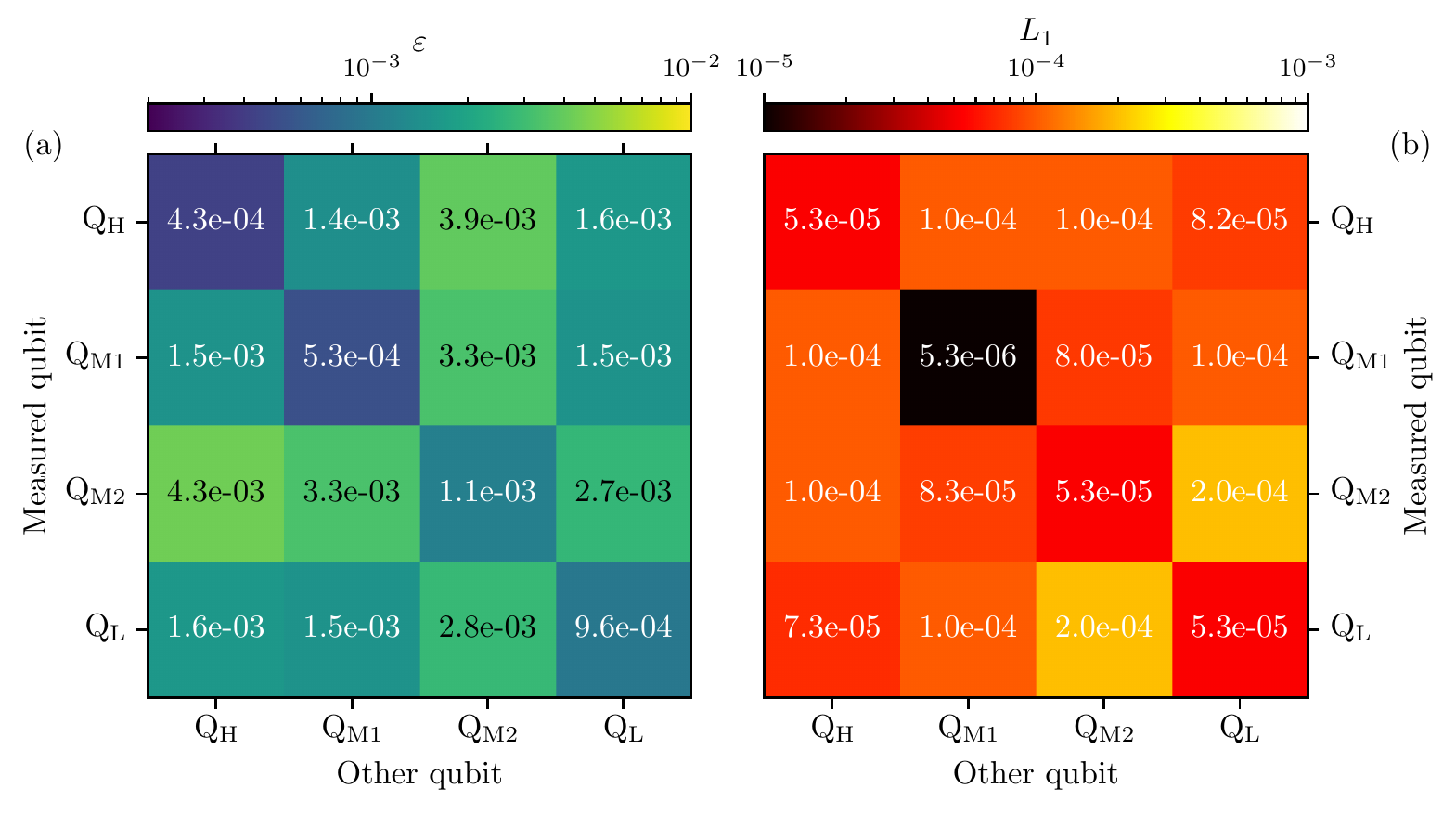}
    \caption{%
    \label{fig:SOM_sqp}%
    Characterization of single-qubit gate infidelity $\Infid$ (a) and leakage $\Lone$ (b) using randomized benchmarking ($100$ randomization seeds). Diagonal elements are extracted from individual single-qubit randomized benchmarking keeping the other 3 qubits in $\ket{0}$. Off-diagonal elements are extracted from simultaneous one-qubit randomized benchmarking on pairs of qubits, keeping the other two qubits in $\ket{0}$.
    }
\end{figure}

\section{Flickering two-level system}
\label{sec:four}
As mentioned in the main text, we were unable to realize the SNZ CZ gate between pair $\QL$-$\QMone$ using the $\ket{11}$-$\ket{02}$ interaction due to the presence of a two-level system (TLS) interacting intermittently with $\QMone$ at the flux amplitude placing $\ket{11}$ and $\ket{02}$ on resonance. Figures~S4(a,b) show the negative impact of this TLS when attempting to characterize the $\ket{11}$-$\ket{02}$ interaction by the standard time-domain chevron measurement. While experience shows that it is probable that such a TLS could be displaced or eliminated by thermal cycling at least above the critical temperature of aluminum, we chose instead to use the more flux distant $\ket{11}$-$\ket{20}$ interaction to realize the SNZ CZ gate for this pair. For this interaction, a standard, stable chevron pattern is observed [Figs.~S4(c,d)].

\begin{figure}[h!]
    \includegraphics[width=0.5\columnwidth]{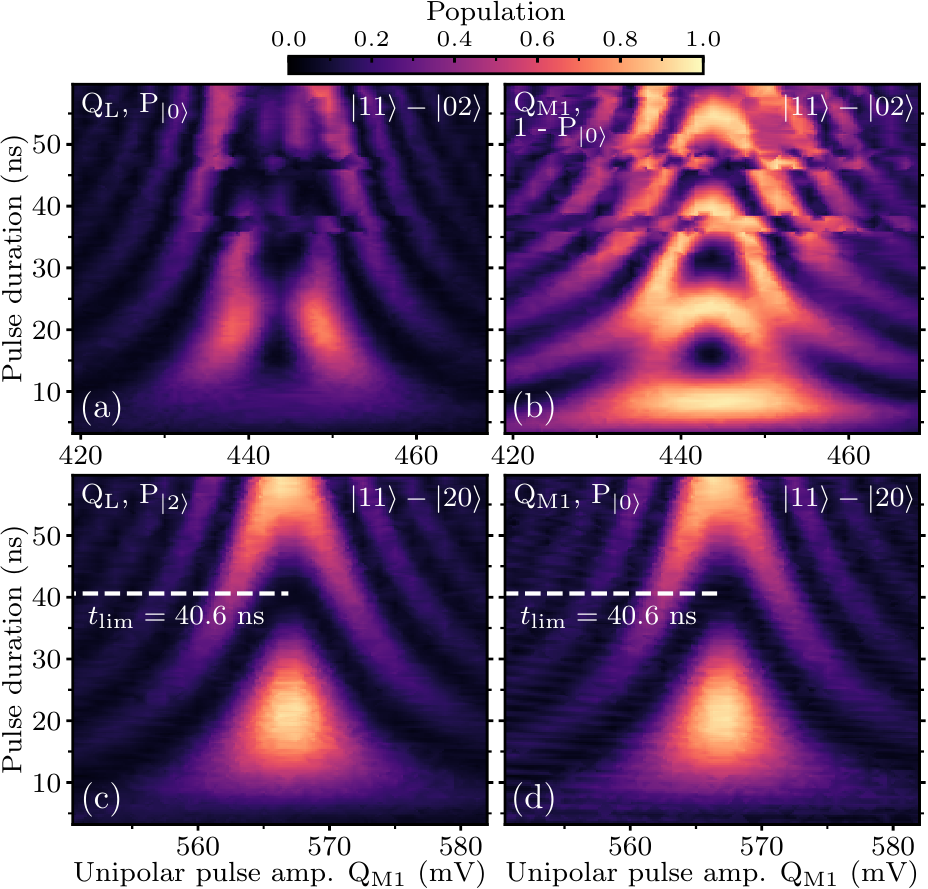}
    \caption{
    \label{fig:SOM_TLS}
    Time-domain characterization of the $\ket{11}$-$\ket{02}$ and (c,d) $\ket{11}$-$\ket{20}$ interactions for pair $\QL$-$\QMone$. (a,b) Landscapes of (a) ground-state population $P_{\ket{0}}$ of $\QL$ and (b) total excited-state population $1-P_{\ket{0}}$ of $\QMone$ as a function of the amplitude and duration of a unipolar square pulse near the $\ket{11}$-$\ket{02}$ resonance. The absence of the expected chevron pattern in these landscapes reflects a flickering TLS resonant with the qubit transition of $\QMone$ at this pulse amplitude.  Horizontally shifting fringes in (a) and (b) are due to flickering of the TLS on the scale of a few minutes. These observations preclude the use of the $\ket{11}$-$\ket{02}$ interaction to realize the CZ gate. In contrast, the landscapes of (c) two-state population $P_{\ket{2}}$ of $\QL$ and (d) $P_{\ket{0}}$ of $\QMone$ and as a function of unipolar square pulse parameters near the $\ket{11}$-$\ket{20}$ resonance reveal a standard, stable chevron pattern. All landscapes were sampled using an adaptive algorithm based on~\cite{Nijholt2019}.
    }
\end{figure}

\section{Residual ZZ coupling at bias point}
\label{sec:five}
Coupling between nearest-neighbor transmons in our device is realized using dedicated coupling bus resonators. The non-tunability of said couplers leads to residual $ZZ$ coupling between the transmons at the bias point. We quantify the residual $ZZ$ coupling between every pair of qubits as the shift in frequency of one qubit when the state of the other changes  from $\ket{0}$ to $\ket{1}$. We extract this frequency shift using a simple time-domain measurement: we perform a standard echo experiment on one qubit (the echo qubit), but add a $\pi$ pulse on the other qubit (control qubit) halfway through the free-evolution period simultaneous with the refocusing $\pi$ pulse on the echo qubit. The results are presented as a matrix in Fig.~S5. We observe that the residual $ZZ$ coupling is highest between $\QH$ and the mid-frequency qubits $\QMone$ and $\QMtwo$.  This is consistent with the higher (lower) absolute detuning between $\QH$ ($\QL$) and the mid-frequency transmons, and the higher (lower) transverse coupling $\Jtwo=\pi/\tlim$ for the upper (lower) pairs.

\begin{figure}[h!]
    \includegraphics[width=0.5\columnwidth]{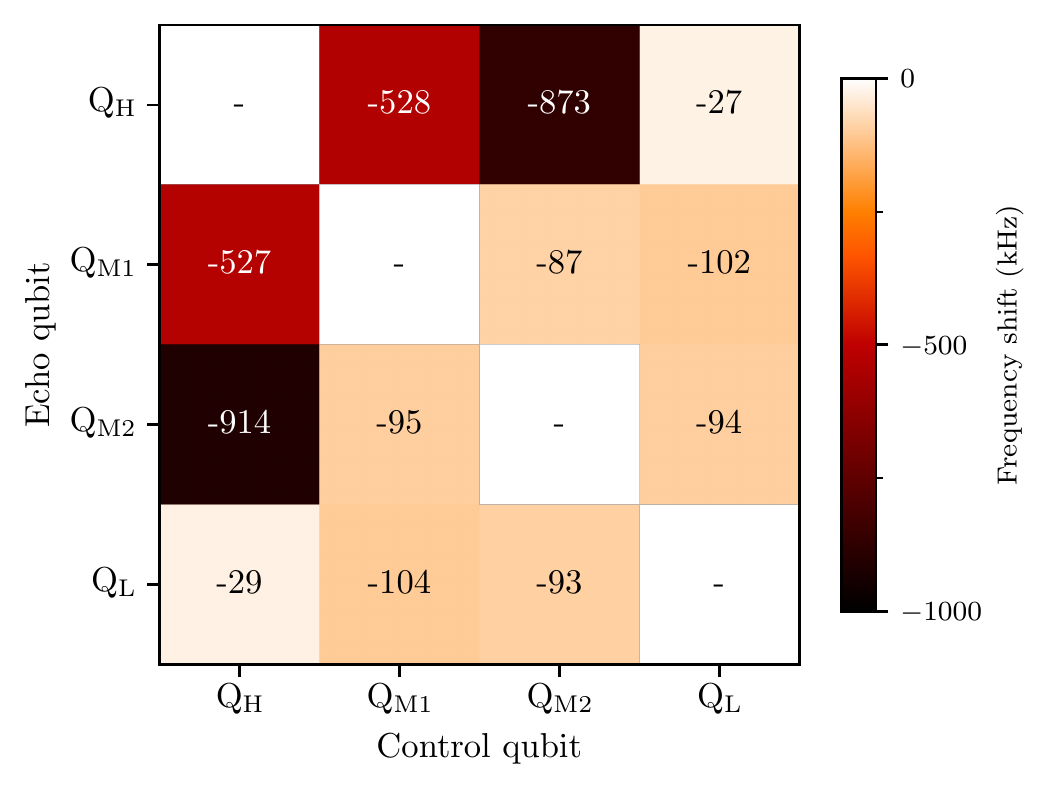}
    \caption{
    \label{fig:SOM_ct}%
    Extracted residual $ZZ$ coupling between all pairs of qubits at their bias points. We report the frequency shift in one qubit (named echo qubit) when the computational state of another qubit (named control qubit) is shifted from $\ket{0}$ to $\ket{1}$.
    }
\end{figure}

An alternative way to evidence this residual $ZZ$ coupling is to extract the fidelity of idling using 2QIRB and to compare this fidelity to that of CZ. To this end, we perform 2QIRB of idling (for $60~\ns$) on pairs $\QMtwo$-$\QH$ and $\QL$-$\QMtwo$. The results, shown in Fig.~S6, show striking differences for the two pairs. For $\QMtwo$-$\QH$, the pair with strongest residual coupling, the idling fidelity is significantly lower than the CZ fidelity. This is because the residual $ZZ$ coupling is a source of error during idling but is absorbed into the tuneup of SNZ. For $\QL$-$\QMtwo$, for which the residual coupling is one order of magnitude lower, this trend is not observed.

\begin{figure}[h!]
    \includegraphics[width=\linewidth]{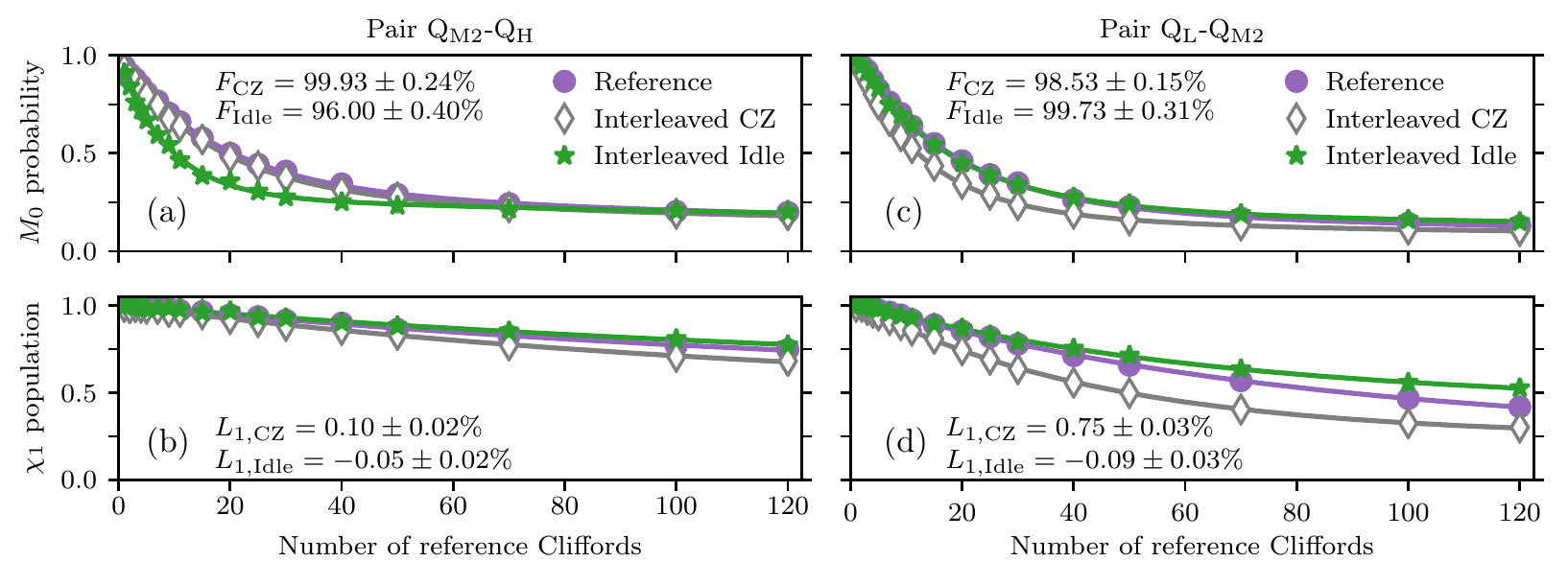}
    \caption{
    \label{fig:SOM_RB_Idle}
    Comparison by 2QIRB of the fidelity and leakage of SNZ CZ versus idling (for an equivalent $60~\ns$) for pairs $\QMtwo$-$\QH$ and $\QL$-$\QMtwo$. SNZ gate parameters are provided in Table~1 of the main text. (a,c) Return probability to $\ket{00}$ as a function of the number $\Ntwoq$ of two-qubit Clifford operations in the reference curve. For $\QMtwo$-$\QH$, the extracted idling fidelity is significantly lower than the SNZ CZ fidelity. This is due to the high residual $ZZ$ coupling between these two qubits as reported in Fig.~S5, which is not refocused during idling but absorbed into the tuneup of the SNZ CZ gate. For $\QL$-$\QMtwo$, idling fidelity exceeds SNZ CZ fidelity as the residual coupling is one order of magnitude weaker. (b,d) Population in the computational subspace as a function of $\Ntwoq$. Leakage as a function of $\Ntwoq$ is weakest when interleaving idling steps, leading to negative $\Li$. This is due to seepage (during idling) of the leakage produced by the reference two-qubit Cliffords.
    }
\end{figure}

\section{Technical details on 2QIRB}
\label{sec:six}

Table~S2 details technical aspects of the characterization of CZ gates by repeated 2QIRB runs.

\begin{table}[h!]
    \begin{tabular}{lcccc}
    \hline
    Parameter & $\QMone$-$\QH$  & $\QMtwo$-$\QH$   &  $\QL$-$\QMone$   &  $\QL$-$\QMtwo$  \\
    \hline
    Number of 2QIRB runs & $39$ & $10$ & $88$ & $35$ \\
    Number of randomization seeds & $100$ & $300$ & $100$ & $100$\\
    Same randomization seeds & No & No & Yes & No \\
    Avg. time per 2QIRB run ($\minutes$) & $17$ & $50$ & $9$ & $17$ \\
    Total wall-clock time ($\hours$) & $28.8$ & $16.9$ & $16.7$ & $14.8$ \\
    \hline
    \end{tabular}
    \caption{
    \label{tab:RB_details}
    Technical details of the characterization of CZ gates by repeated 2QIRB.  The average time per 2QIRB run is the time required to perform back-to-back measurements of the reference and the CZ-interleaved curves.
    The total wall-clock time includes the overhead from compilation of RB sequences and other measurements performed in between the CZ 2QIRB runs, e.g., idling 2QIRB (Fig.~S6).
    }
\end{table}

\end{bibunit}
\end{document}